# Amino-acid network clique analysis of protein mutation correlation effects: a case study of lysozme

**Dengming Ming[1], Rui Chen[1] and He Huang[1,2]**

[1]College of Biotechnology and Pharmaceutical Engineering, [2]College of Pharmaceutical Sciences, 30 Puzhu South Road, Nanjing, Jiangsu 211816, PR China

*Contact information: Dengming Ming, Biotech Building Room B1-404, College of Biotechnology and Pharmaceutical Engineering, Nanjing Tech University, 30 South Puzhu Road, Jiangsu 211816, PR China; Tel: 8625-58139942; Email: dming@njtech.edu.cn, biotech@njtech.edu.cn

Abstract: Optimizing amino-acid mutations has been a most challenging task in modern bio-industrial enzyme designing. It is well known that many successful designs often hinge on extensive correlations among mutations at different sites within the enzyme, however, the underpinning mechanism for these correlations is far from clear. Here, we present a topology-based model to quantitively characterize correlation effects between mutations. The method is based on the molecular dynamic simulations and the amino-acid network clique analysis that simply examines if two single mutation sites belong to some 3-clique. We analyzed 13 dual mutations of T4 phage lysozyme and found that the clique-based model successfully distinguishes highly correlated or non-additive double-site mutations from those with less correlation or additive mutations. We also applied the model to the protein Eglin c whose topology is significantly distinct from that of T4 phage lysozyme, and found that the model can, to some extension, still identify non-additive mutations from additive ones. Our calculations showed that mutation correlation effects may heavily depend on topology relationship among mutation sites, which can be quantitatively characterized using amino-acid network k-cliques. We also showed that double-site mutation correlations can be significantly altered by exerting a third mutation, indicating that more detailed physico-chemistry interactions might be considered with the network model for better understanding of the elusive mutation-correlation principle.

---

## 1. INTRODUCTION

Successful protein design in enzyme engineering often hinges on a good understanding of the relationship between protein structures and their biological functions. One of the key steps in rational design is the introduction of special amino-acid replacements at particular sites of the studied proteins, which is expected to enhance the protein thermo-stability, catalytic activity, etc. In practice, effective designs often involves simultaneously mutations at two or more sites in the target proteins. Thus one critical question concerning mutation design arises: is there any correlation between mutations at different sites in the studied proteins? If so, can we predict them? Obviously, if mutations at different sites are independent from one another, then the overall effect of the multiple-mutation can be estimated by summing up the effect of every single mutation and is called to be additive[1]. On the contrary, in cases where strong interplay between mutations at different sites exists, the overall mutation effects are unpredictable from those of single mutations and exhibit non-additive.

Mutation additivity effects had been studied in a variety of backgrounds in early days by many structural biologists. For example, Sandberg and Terwilliger[2] examined the additive effects of mutations in gene V proteins, and found that different types of mutations showed strong additivity. In addition, they found that mutations at sites that have intense van der Waals interactions tend to be weaker additive. Boyer and colleagues[3] suggested that the non-additivity of mutations at distant sites indicates an information communication between amino acids at these sites, and they called “thermodynamic coupling” for the enhanced thermo-stability due to this non-additive phenomenon. They used atomic resolution NMR to examine the hydrogen exchange in the enzyme at its natural state, and had attempted to determine the dynamic perturbation between the two mutation sites. They concluded that thermodynamic coupling between distal sites was caused by physical interactions between amino acids at these sites in the natural structure of the studied protein.

T4 phage lysozyme, as a model protein for studying the relationship between protein structure and their functions, was also one of the early model proteins used for study of mutation correlation effects. Matthews and colleagues[4] observed that mutations that introduce negative charges at ends of alpha helices in T4 phage lysozyme and produced electrostatic interactions at these sites, such as S38D and N144D, were additive. They designed a series of combinatorial mutations at distant sites and do not form direct contact. Interestingly, most of these multiple mutations are found to be very strong additive, and they can form either direct physical contacts or not. An extreme example that used the mutation additive effect is a combination of 7 mutations, S38D/A82P/N144D/I3L/V131A/A41V/N116D, which was found to have the largest melting temperature increase of 8.3°C [5].

One the other hand, some mutations that involves direct physical interactions did exhibit strong non-additivity. For example, the double-site mutant A98V/T152S showed strong non-additivity compared with the corresponding single mutations, melting temperature change caused by the double-site mutation is 7.6°C less than the summation of those imposed by the two corresponding single mutations[6, 7]. An examination of the native structure reveals that A98 and T152 orients to one another in the 3D structure and form direct contact. Matthews and colleagues[8] suggested that the dynamic perturbation a mutation introduces will start at the mutation site and spread to its neighboring sites. The stronger the neighboring structure can absorb the impact caused by a mutation, the smaller the change of thermal stability might be introduced by the mutation. In the case of A98V/T152S, due to the relative strong interaction between the two sites, the capabilities for immediate neighboring structures to absorb impact imposed by mutations at either sites or mutations at both sites are significantly dependent on the detailed interactions: presumably mutations that enhance the contact between two sites might weaken the power for neighboring structures to relieve the mutation perturbations at these sites, thus causing larger thermal stability changes. Other mutations that do not involve direct interactions were also found having strong non-additivity[9].

Undoubtedly, it is import to understand the mechanism underpinning the mutation correlation effects, and predetermining mutation additivity at selected sites can effectively reduce experimental workload in rationale design. Recently, as the accumulation of mutation data methods had been developed for prediction of mutation effects. For example, Tian etc. [10] developed a machine-learning method to predict mutation effects on protein thermo-stability based on a 3366 mutant protein database. Pires etc.[11] predicted missense mutations using some graph-based signatures. Very recently, Dehghanpoor etc. [12] compared the performances of several machine learning methods. However, up-to-now accurate prediction of mutation correlation effect can be still a very challenging task.

In this paper, a mathematical model based on protein structural amino-acid network was presented that successfully isolate double-site mutations with strong non-additivity from additive ones for T4 bacteriophage lysozyme. We studied different factors in the protein topological network that show high correlation with the mutation additivity. Double-site mutations of T4 phage lysozyme were selected if the two corresponding single-site mutations were also measured, and the correlation effect of mutations at the two sites were determined based on the measured thermodynamics data[9]. The dependence of mutation correlation effect on the distance between the two sites was examined. We then presented a protein topological network model based on amino acid interaction[13], and examined the network topological quantities and their relationships with the mutation correlation effects. Finally, we derived a mathematical model based on protein network topology to predict mutation additivity/non-additivity. The model was also successfully applied to a new protein Eglin c whose structural has a distinct topology from that of T4 phage lysozyme.

## 2. Method

*2.1. Preprocessing the experimental mutation data and selection.*

The T4 bacteriophage lysozyme mutation data are taken from reference [9]. We first ignored those data that lacking thermodynamic measurement or melting temperature changes. Then, the experimental data of double-site mutants were examined, and those were kept for further analysis whose two corresponding single-site mutations that makes up the double-site mutation was present. The additive effect of a boule mutation was measured based on the thermodynamic quantities as follows:

$$\Delta\Delta G_{sum} = \Delta\Delta G_i + \Delta\Delta G_j \tag{1a}$$

$$\Delta\Delta\Delta G_{ij} = \Delta\Delta G_{ij} - \Delta\Delta G_{sum} \tag{1b}$$

where $\Delta\Delta G_i$, $\Delta\Delta G_j$ are the Gibbs free energy changes due to the single point mutation at site *i* and *j*, and $\Delta\Delta G_{ij}$ is that due to the double-site mutations at both sites $i$ and $j$. $\Delta\Delta G_{sum}$ evaluates the total effect due to the two single-site mutations in an ideal case when the two mutations are completely independent. $\Delta\Delta\Delta G_{ij}$ measured the difference between the observed double-site mutation effect and the ideal effect when the corresponding two single-site mutations are additive. In other words, $\Delta\Delta\Delta G_{ij}$ reveals how far a double-site mutation deviates from a perfect additive one. In this sense, the larger the absolute value of $\Delta\Delta\Delta G_{ij}$, the less likely the studied double-site mutation being additive and having a larger chance to be non-additive. To examine the possible dependence of double-site mutation correlation effect on the distance between the two involved sites, we defined the distance as the length of a virtual edge linking the two $C_\alpha$ atoms in the wild-type lysozyme structure (PDB code 2LZM[14]).

*2.2. The equilibrium dynamics conformation ensemble.*

The 3D structure deposited in PDB usually captures the frozen snapshot of protein in typical crystal-packing state, which might be significantly different from its functioning conformations. Here, we derived a series of conformation ensembles using the conventional molecular dynamics simulation techniques. All simulations were performed using the simulation package of GROMACS(version 4.5.4) [15] and the Oplsaa all atom field[16], with the lysozyme placed in a cubic water box. The starting lysozyme conformations with different mutations were taken from X-ray structures whose PDB entry codes are listed in reference[9]. For mutant structures whose structures were not solved and not available in PDB, we built their structural models based on the that of wild-type protein (PDB code 2LZM[14]) using the homology modeling program of MODELLER version 9.4[17]. All simulations were carried out in a temperature of 320 K, a pressure of 1 atm, a time step of 2 fs and a non-bond cutoff of 12Å. For each simulation system, certain number of Na+ ions were added to neutralize the system and a layer of water molecules with thickness of about 1 nm was added to the solvate the proteins. The PME (Particle, Mesh, Ewald, PME) algorithm was applied in calculating the long-range electrostatic interactions. After an initial minimization and a 1ns steric relief equilibrium simulation, each system was then performed a total 100ns productive simulation. We collected the snapshots of each system every 10 ps, and recorded a total of 10 thousand conformations for each mutants which were used for further analyzes.

*2.3. The amino-acid interaction network*

To understand the mutation correlation effect, we examined the topologies of the studied protein structures and focused on the amino acid networks. The amino-acid networks had been used in studying different biophysical problems such as the protein folding, catalysis, as well as the mutation perturbations [18-21]. We built the network based on the amino acid interactions, which were determined using the program RING-2.0 (Residue Interaction Network Generator[13]). The program determined the most common types of physicochemical interactions that are indispensable in maintaining the protein 3D structure, including hydrogen bonds, disulfide bonds, Van der Waals interaction, electrostatic interaction, π-π stacking interactions, π-cation interactions. Table 1 listed typical parameters in determining the interactions and energetics. The network was created by using alpha carbon atoms as nodes, and the edges were generated between two neighboring nodes whose amino acids were found to form direct interactions. Thus, in such a network most amino acids are connecting to one another, and two amino acids are found either directly linked through an edge or indirectly connected via some intermediate linkers. In some cases, there are also scattered a few isolated nodes where the amino acids have non connection with any surrounding residues. The distance between two given amino acids was counted as the number of edges in the shortest path linking the two nodes within the network.

**Table 1.** The type of interaction bonds and their energetic parameters used in defining protein amino-acid interaction network

| Bonds | Cutoff （Å） | Energy (KJ/mol) |
|---|---|---|
| Hydrogen bond | 3.5 | 17.0 / 40.0 / 115.0 (on distance) |
| Van der Walls | 0.5 | 6.0 |
| Disulfide bond | 2.5 | 167.0 |
| electrostatics | 4.0 | 20.0 |
| π-π stacking | 6.5 | 9.4 |
| π-cation | 5.0 | 9.6 |

Note: the distance in hydrogen bonds refers to that between the hydrogen donor and acceptor atoms. The distance of van der Waals interaction is that between the surfaces of two atoms. The distance in sulfur bonds refers to that between the two sulfur atoms. The distance used in electrostatic interaction calculations are measured between mass centers of the two oppositely charged groups. The distance in π-π stacking interaction refers to those of the geometric centers of benzene rings of the aromatic residues. The distance in a π-cation interaction is measured from the mass center of the positively charged group in a residue to that of the benzene ring in another residue. The energy of the action is averaged over the various cases of the same type of interaction, which is a rough approximation of the corresponding real interaction.

*2.4. The k-clique community in the amino-acid network*

One interesting topic concerning amino-acid networks is to examine the geometric pattern emerged from protein structural topology and to analyze their meaning in the sense of biological function. We analyzed the network pattern using the Networkx package version 1.11 [22] developed for programing language Python version 3.6. A network can be divided into a few domains, and nodes inside a domain tend to form dense connections among one another while those belonging to different domains show very weak connections. The nodes in a domain build up the so-called community, which is further divided into a series of connected subgraphs, called the k-cliques, using a clique percolation method[23, 24].

Specially, a *k*-clique is a complete subgraph of *k* nodes in which each pair of nodes is connected by an edge, indicating a strong and intense mutual interactions happen among amino-acids on these nodes. Two cliques are regarded as to be adjacent if there are *k*-1 edges linking the two cliques. Further, two-cliques are regarded as to be interconnected, if one can find a way to connect one *k*-clique to the second one through several intermediate adjacent *k*-cliques. A collection of all interconnected *k*-cliques in a given network defines a *k*-clique community. In this sense, a network can be simplified by dividing into a few *k*-clique communities. There may be nodes that belong to different clique communities that are not connecting with each other. Considering the dynamic feature of the protein structures, it is likely *k*-clique community distribution of amino-acid network may be perturbed to some extent due to thermodynamic fluctuations. Thus, we calculated the ensemble of *k*-clique communities for each mutant structures derived from the molecular dynamic simulations, and to examine the mutation correlation effects we compared *k*-clique community distributions and their changes upon different mutations. For simplicity, in the reset of the paper a *k*-clique is simply referred to as a *k*-clique community.

*2.5. The 3-clique and correlation effects in double-site mutations*

Considering that amino-acid within the same clique tends to have tight connection than does amino-acid belonging to different cliques, it is interesting to examine whether or not mutations with higher correlation effect tends to locate in the same clique. Specifically, for a given double-site mutation we calculated the probability by which the two mutation sites belong to the same 3-clique and the correlation effect involving the double-site mutations as follows:

1. Generating an ensemble of protein conformation from an 100ns equilibrium dynamics simulation of the studied enzyme. The combination of ten thousands of snapshots in aqueous solution should be better representation of the interactions within the protein in functioning conditions.
2. Determining 3-cliques for network of each snapshot using the Networkx package [22].
3. Calculating the proportion $P_{ab}$ of snapshots in the ensemble in which the two sites (*a*, *b*) of the studied double-site mutation belong to a 3-clique:

$$P_{ab} = \frac{\sum_{i=1}^{N} C_{ab}(i)}{N} \tag{2a}$$

where *N* represents the total number of equilibrium snapshots, here it equals to $10^4$, and

$$C_{ab}(i) = \begin{cases} 1 & a \text{ and } b \text{ in a 3-clique} \\ 0 & \text{else} \end{cases} \tag{2b}$$

$P_{ab}$ measures the probability that two sites (*a*, *b*) are kept in some 3-clique due to either direct or indirect interactions among amino-acid interactions. The closer $P_{ab}$ is to 1, the more likely *a* and *b* tends to have a tight connection from the topological perspective.

*2.6. The validation models*

To evaluate the relationship between clique-probability $P_{ab}$ and the correlation effects of a double-site (*a*, *b*) mutation which is quantified by the value of addivity, we first investigated the addivities of the 13 double-site mutations of the T4 phage lysozyme, and then determined $P_{ab}$ for each of them from the simulations and network modeling. We also examined the eglin c protein[3] who has a distinct structural topology with that of T4 phage lysozyme. Finally, we also studied double-site mutations in which the two involved sites are far from each other but have a high $P_{ab}$ value, which, according to our prediction, might have a high probability to have non-additivity effect.

## 3. Results and Discussion

*3.1. The correlation effects in double-site mutations is independent of the mutation-site distance*

Figure 1 shows that most of the double-site mutations are strong additive, and the two double-site mutations, the S117I/N132I and A98V/T152S, show significant non-additivity. Interestingly, compared with its component single mutations the non-additive effect of the doubsite-site mutant A98V/T152S significantly decreases the enzyme thermo-stability due to increment in free energy change, whereas correlation effect in S117I/N132I makes the enzyme stable due to a decrement in free energy change. Quantitatively, non-additive effect of mutant S117I/N132I is weaker than that of A98V/T152S. There is no obvious correlation between site distance and the additivity effect of double-site mutation (see Table 2 and Figure 2). We noticed that the distances between the sites of double-site mutations with strong non-additivity are relatively small, and the reverse scenario is not necessarily true, i.e., double-site mutations exhibit very weak non-additivity whose mutation-site-distance are actually very short. No strong non-additivity effect was observed for long-distance double-site mutant. Presumably, intensive interference between the two sites of a double-site mutation might be required in order to exhibit a strong non-additivity effect, and this interferential interaction might be lack or very weak when the two sites are well-separated. However, for relative short double-site mutation, it is still interesting to understand why a few of them, such as E128A/V131A, are non-additive while a majority of them are still additive.

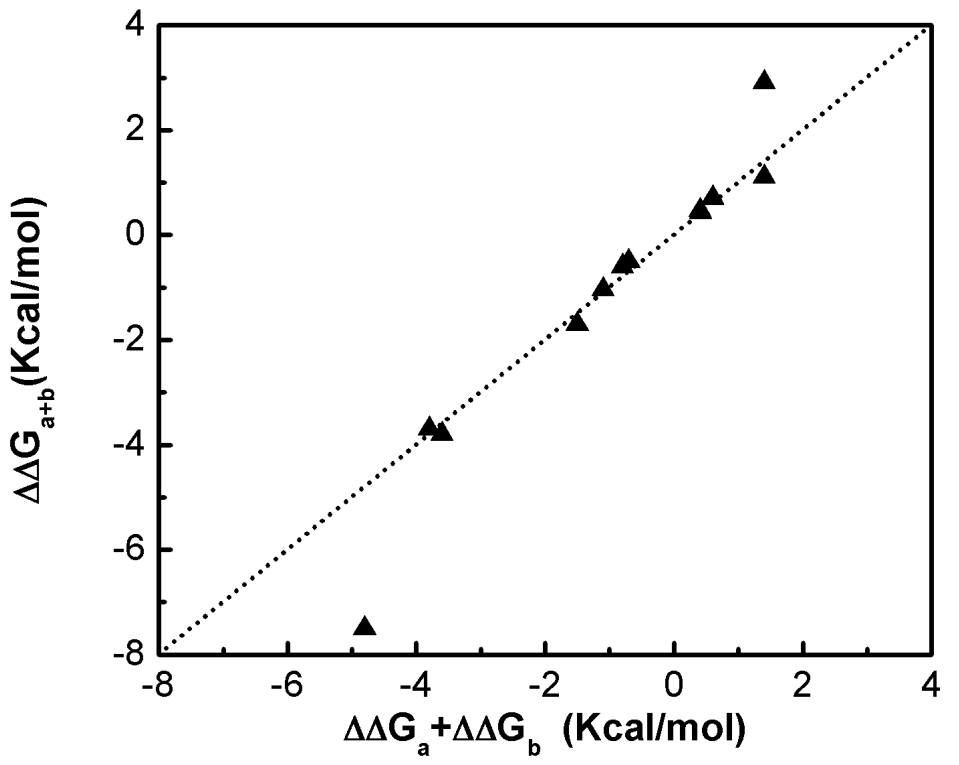

**Figure 1.** Compare the free energy changes of double-site mutations with that derived by the summation of the two corresponding single-site mutations of the T4 phage lysozyme.

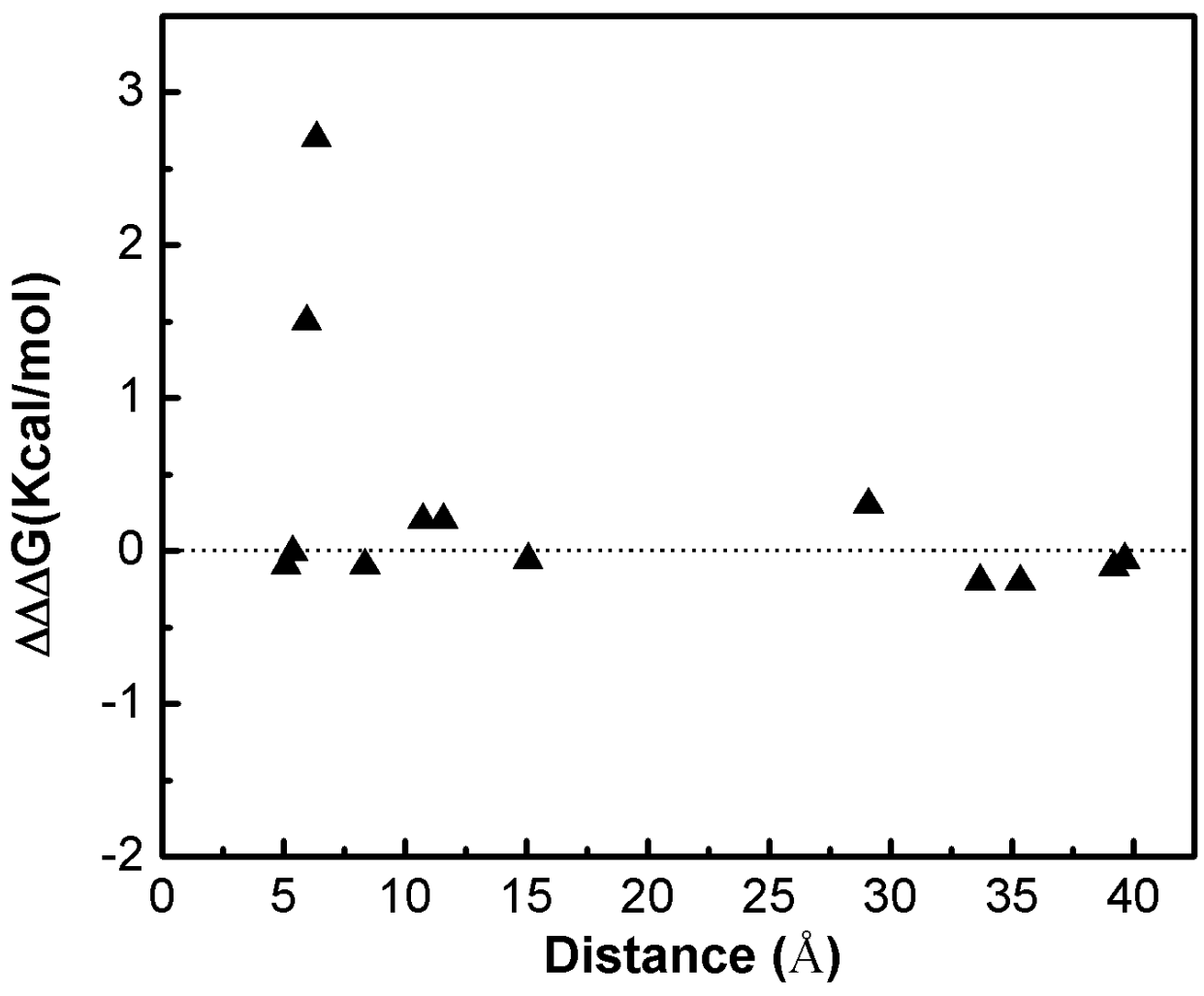

**Figure 2.** The dependence of the double-site mutation effects on the distances between the two corresponding single sites of the T4 phage lysozyme.

**Table 2.** The 13 double mutations in T4 phage lysozyme.

| Mutations | ΔΔΔG (kcal/mol) | $C_\alpha$ atom distance(Å) |
|---|---|---|
| S117I/N132I | -1.5 | 5.95 |
| K16E/K135E | -0.2 | 33.69 |
| K16E/R154E | -0.2 | 35.33 |
| A41V/V131A | -0.11 | 39.20 |
| N116D/R119M | -0.1 | 5.09 |
| D89A/R96H | -0.1 | 8.35 |
| R119E/K135E | -0.06 | 15.06 |
| K16E/R119E | -0.06 | 39.63 |
| E128A/V131A | -0.01 | 5.38 |
| K135E/K147E | 0.2 | 11.58 |
| K85A/R96H | 0.2 | 10.73 |
| S38D/N144D | 0.3 | 29.07 |
| A98V/T152S | 2.7 | 6.35 |

**Table 3.** Typical 3-cliques found in a T4 phage lysozyme amino-acid network.

| Clique | Nodes (residue index) |
|---|---|
| 1 | 67, 4, 101, 70, 7, 104, 71, 11 |
| 2 | 34, 38, 42, 25 |
| 3 | 66, 46, 50, 54, 58, 27, 62 |
| 4 | 66, 70, 31 |
| 5 | 96, 99, 100, 103, 75, 78, 88, 91 |
| 6 | 91, 126, 95 |
| 7 | **98**, **152**, 156, 94 |
| 8 | 101, 105, 145, 149 |
| 9 | 128, 129, **132**, **117**, 120, 125 |
| 10 | 138, 139, 142, 146, 149, 150 |
| 11 | 160, 148, 151 |

*3.2. Double-site mutations that have strong correlation effects tend to have their sites located within a 3-clique and that of weak correlation have their sites located in different 3-cliques.*

The calculated 3-cliques of T4 bacteriophage lysozyme vary in size and locations (Table 4). It is interesting to examine mutation effects at every site in each clique and the additivity properties between these sites within each clique. For this sake, we first consider the additive double-site mutants, and found that for each such mutant whose mutated two sites do not belong to any 3-clique. We then checked the site distribution for all the non-additive double-site mutants, and found that the two mutation sites of each such mutant can be identified in some 3-clique community (Cliques 7 and 9). Spatial arrangements of clique members in lysozyme network (Figure 3) indicates that 3-cliques have relative uniform distribution, while larger cliques tend to form at linker area that joins the two lobs of the enzyme.

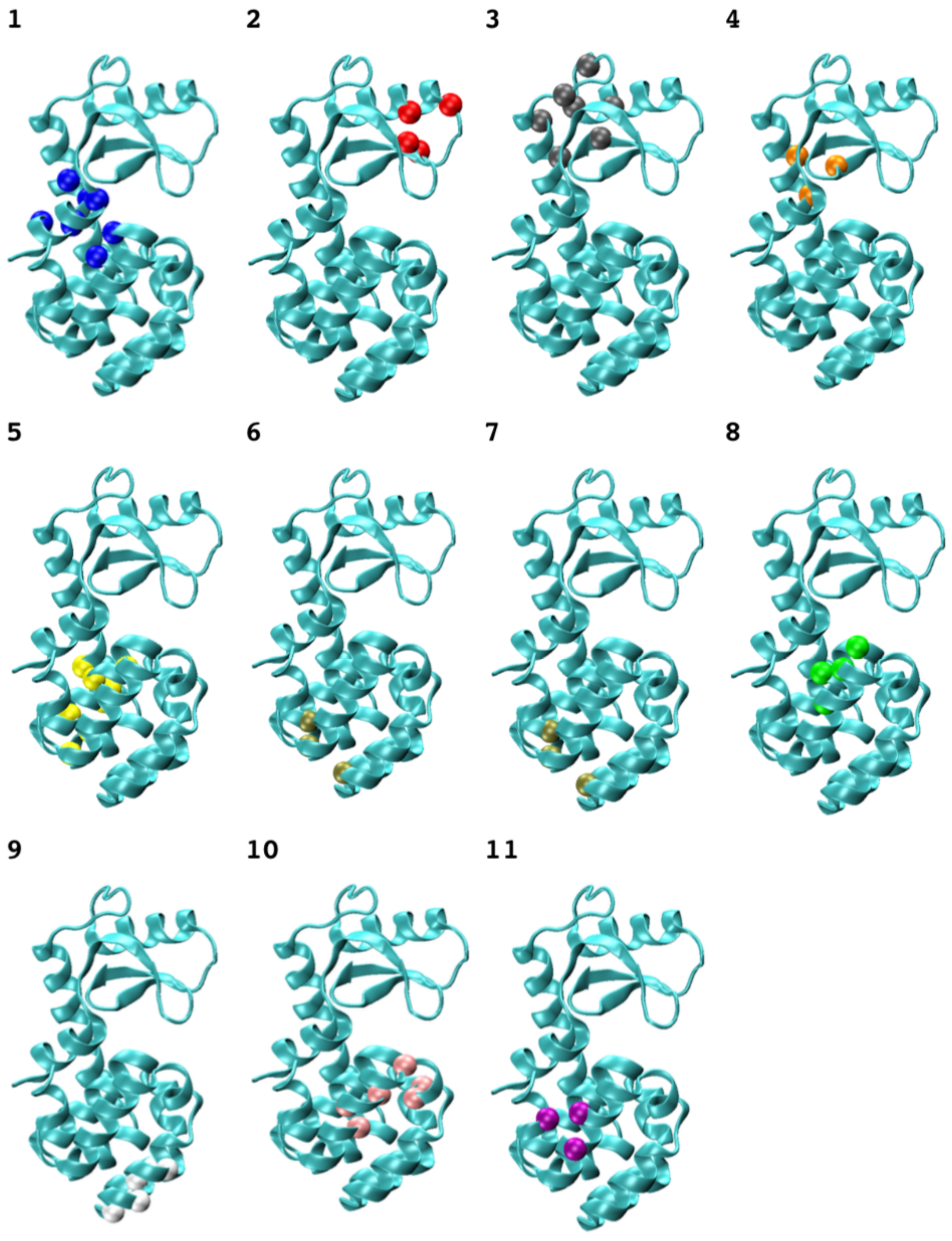

**Figure 3.** The typical 3-cliques of an amino-acid network of T4 phage lysozyme. The clique nodes are represented by alpha-carbon as shown in ball, and the labels correspond to the factions listed in Table-4.

**Table 4.** $P_{ab}$ for double-site mutations derived from different T4 phage lysozyme models. WT stands for wild type lysozyme structure (PDB code 2LZM[14]), K16E for the mutant structure (PDB code 1L42[25]), R154E for the mutant structure (PDB code 1L47[25]), the four structures of K16E/R154E, S117I, N132I, S117I/N132I are homology models derived by MODELLER[17] based on the structure of wild type lysozyme with the corresponding mutations of K16E/R154E, S117I, N132I, S117I/N132I.

| Double mutations | Models | | | | | | |
|---|---|---|---|---|---|---|---|
| | WT | K16E | R154E | K16E/R154E | S117I | N132I | S117I/N132I |
| (116,119) | 0 | 0 | 0 | 0 | 0 | 0 | 0 |
| (117, 132) | 0.48 | 0.45 | 0.28 | 0.56 | 0.60 | 0.39 | 0.52 |
| (119, 135) | 0 | 0 | 0 | 0 | 0 | 0 | 0 |
| (128, 131) | 0 | 0 | 0.01 | 0 | 0 | 0 | 0 |
| (135, 147) | 0 | 0 | 0 | 0 | 0 | 0 | 0 |
| (16, 119) | 0 | 0 | 0 | 0 | 0 | 0 | 0 |
| (16, 135) | 0 | 0 | 0 | 0 | 0 | 0 | 0 |
| (16, 154) | 0 | 0 | 0 | 0 | 0 | 0 | 0 |
| (38, 144) | 0 | 0 | 0 | 0 | 0 | 0 | 0 |
| (41, 131) | 0 | 0 | 0 | 0 | 0 | 0 | 0 |
| (85, 96) | 0.07 | 0.05 | 0.05 | 0.05 | 0.12 | 0.11 | 0 |
| (89, 96) | 0 | 0 | 0 | 0 | 0 | 0 | 0 |
| (98, 152) | 0.43 | 0.40 | 0.24 | 0.03 | 0.02 | 0.56 | 0.12 |

$P_{ab}$ of the additive double-site mutations were determined to be 0 (see Table 4), indicating that the probability of these mutations in the wild-type T4 bacteriophage lysozyme was 0 in the presence of 3-cliques. Therefore, the results of model calculations suggest that there is no additive effect in the combination of these mutations. This is consistent with the experiment obervation[1]. Although $P_{85,96}$ is not zero it is so small that this double-site mutation can hardly be assumed to be non additive. $P_{ab}$ of non-additive double-site mutations are remarkably different from those of additive double-site mutations, having a value varied mostly from 0.4 to 0.5 depending on structural models of the enzyme. Essentially, the $P_{ab}$ values calculated based on wild-type structure are representative to measure the correlation effects of the double-site mutations in lysozyme. These results suggest that the additive effects of double-site mutation can be closely related to the topological feature of protein amino-acid network and less dependent on the detailed pysico-chemical interactions involved inside the protein.

It is interesting to notice that $P_{ab}$ values do vary when measured with different mutant structures. For example, while $P_{85,96}$ for the non-additive dual-mutation at sites 85 and 96 was determined close to 0 in three mutants K16E, R154E, and K16E/R154E, it reaches 0.1 in both mutants S117I and N132I, and, at the same time, this value decays to zero again in dual-mutant S117I/N132I. Another case happens in A98V/T152S, while $P_{98,152}$ has significant value for most of the examined structures, it almost drops to zero in mutants K16E/R154E and S117I. These results suggest that the correlation effects of a double-site mutation might be affected by a third (or a forth) mutation.

### 3.3. *Eglin c*

We examined the 3-cliques in the amino-acid network of a new protein Englic c whose topology is distinct from that of lysozyme (Table 5). It is interesting to noticed that compared with the two zero probability double-site mutations V18I/L27I and V34L/P58Y, the non-additive mutation V18A/V54A does show non-zero probability, indicating that the 3-clique relationship between the two mutation sites may still play a role in distinguishing non-additive mutations from additive one. However, the relative small value of $P_{18,54}$ suggests that some information is still missing to fully isolate this non-additive dual mutation from the other two, which is necessary for well explaining the correlation effects of the examined mutation sites.

**Table 5.** $P_{ab}$ for double-site mutations for Eglin c protein. The simulation is based on wild type Eglin c (PDB code 1EGL[26]), and the distance is measured between the two mutation sites.

| Double mutation | Distance (Å) | ΔΔΔG（kJ/mol） | $P_{ab}$ |
|---|---|---|---|
| （18,27） | 14.2 | -0.12±0.20 | 0 |
| （18,54） | 8.2 | -0.95±0.19 | 0.11 |
| （34,58） | 17.2 | -0.38±0.13 | 0 |

**4. Conclusions**

Protein mutation effects have been becoming a popular topic in cell biology due to recently developed deep scanning technique, which creates large-scale mutagenesis data that associates intrinsic protein structures and functions with the consequences of relevant genetic variation[27]. A critical question arise from this scenario is that how natural selection works with the innumerable yet almost random mutations in the so-called evolution process? In this paper, we examined possible intrinsic correlations between protein random mutations based on protein structural network calculations. We analyzed the additivity effects of 13 double-site mutations of T4 bacteriophage lysozyme, and found that mutations at distal sites are usually strong additive while those occurred at neighboring sites can be either additive or non-additive. To systematically estimate correlation effects of double-site mutations, we investigated the amino-acid network structures for each mutant, and determined the topological quantities of these networks. We generated equilibrium configuration ensembles of the studied proteins using conventional simulations and built the amino-acid network for each structure. We then analyzed the topological characteristics of the protein networks, such as the distribution of k-cliques, and found significant correlation between 3-faction associations and the double-site mutant additivity: non-additive mutations tend to happened between sites belong to the same 3-clique structure. It was found that the clique model could significantly separate non-additive double-site mutations from those additive ones for the examined proteins. Our calculations also suggested that such correlation probabilities can be changed to some extension by applying a third mutation.

Although the faction group model used here is very simple it does work very well for lysozyme structures. We also noticed that the model cannot explain mutation correlation effects for some different proteins such as myoglobin[28]. Another weak point with the model is that it tends to create very few 3-cliques for many proteins, especially for those protein whose network topology are relatively sparse, which usually resulted in false negative predictions. It becomes even more complicated when considering the perturbation due to a third mutation. Thus, we expect to refine the models in the near future by combining the simple network analysis as shown in this work with the detailed physico-chemistry characterization and give fruitful understanding of protein mutation effects.

Acknowledgements: One of the author (DM) thanks Hong Qian in University of Seattle for his helpful discussion on the lysozyme mutation data. This work was supported, in part, by the National Key Research and Development Program of China (Grant No. 2017YFC1600900) and by the Key University Science Research Project of Jiangsu Province (Grant No. 17KJA180005).